# Journal of Physics Communications

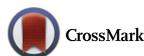

**PAPER**

**OPEN ACCESS**

# Inverse spectrum problem for quasi-stationary states

Sebastian H Völkel 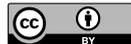

Theoretical Astrophysics, IAAT, University of Tübingen, Germany

E-mail: sebastian.voelkel@uni-tuebingen.de

## Abstract

In this work we present a semi-classical approach to solve the inverse spectrum problem for one-dimensional wave equations for a specific class of potentials that admits quasi-stationary states. We show how inverse methods for potential wells and potential barriers can be generalized to reconstruct significant parts for the combined potentials. For the reconstruction one assumes the knowledge of the complex valued spectrum and uses the exponential smallness of its imaginary part. Analytic spectra are studied and a recent application of the method in the literature for gravitational wave physics is discussed. The method allows for a simple reconstruction of quasi-stationary state potentials from a given spectrum. Thus it might be interesting for different branches of physics and related fields.

## 1. Introduction

The inverse spectrum problem for one-dimensional wave equations with a potential term has a long history and has led to different approaches for solving it, see [1] for an outstanding and classical introduction. In this work we are interested in reconstructing a potential of the kind shown in figure 1. It should be characterized by a potential well with one minimum next to a potential barrier with only one maximum. The presented method is based on the generalization of semi-classical methods for pure potential wells and potential barriers that use the 'inversion' of the Bohr–Sommerfeld rule and Gamow formula, respectively [1–4]. We show how the individual methods can be generalized to our class of potentials and under what conditions a unique reconstruction is possible.

The purpose of this work is to present a self-contained description of an inverse method, first developed for the study of ultra compact stars in general relativity [5]. Such exotic compact objects have recently drawn a lot of attention in the field, since they could potentially mimic black holes and show quantum gravitational effects on the horizon scale. For a current review of the field we refer to [6]. However, the method is rather general and also applicable in other fields, thus we want to present and demonstrate it to a broader audience here. After reviewing the necessary ingredients, we apply the method for the first time to analytic functions that serve as candidates for quasi-stationary state spectra and discuss the potentials one can reconstruct from it.

This paper is organized as follows. We will review the classical Bohr–Sommerfeld rule and its generalization to quasi-stationary states in section 2. The subsequent section 3 presents the relevant results for the inverse problem of potential wells and potential barriers, as well as our generalization to quasi-stationary sates. Analytic results of the method for different spectra are derived and discussed in section 4. There we also summarize the recent application of the method in the theoretical study of gravitational waves from ultra compact objects [5]. We discuss our findings in section 5. The conclusions are presented in section 6. Throughout the paper we set $\hbar = 2m = 1$.

## 2. Bohr–Sommerfeld rules

Before solving the inverse problem, let us recall the 'direct' problem first. By this we mean to obtain the spectrum $E_n$ for a given potential $V(x)$ that appears in the one-dimensional wave equation

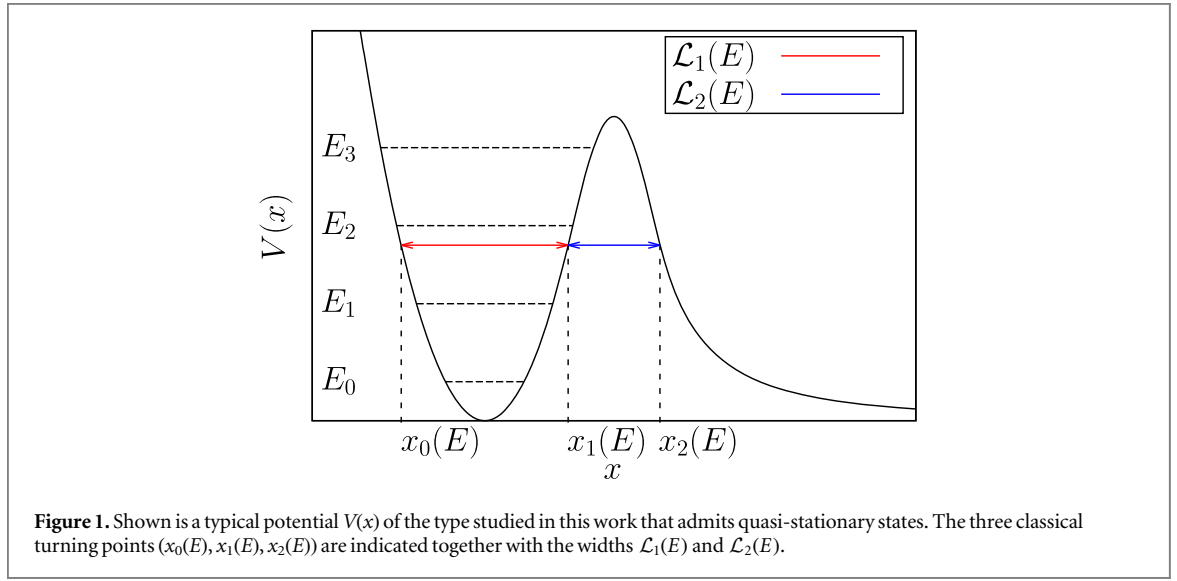

**Figure 1.** Shown is a typical potential $V(x)$ of the type studied in this work that admits quasi-stationary states. The three classical turning points $(x_0(E), x_1(E), x_2(E))$ are indicated together with the widths $\mathcal{L}_1(E)$ and $\mathcal{L}_2(E)$.

$$\frac{d^2}{dx^2}\Psi(x) + (E_n - V(x))\Psi(x) = 0. \tag{1}$$

A simple method to get an approximative solution for the spectrum $E_n$, if $V(x)$ is a potential well, is the Bohr–Sommerfeld rule shown in equation (2). It can be derived with the Wentzel-Kramers-Brillouin (WKB) method [7] and yields an approximative solution following from the integral equation

$$\int_{x_0}^{x_1} \sqrt{E_n - V(x)}\, dx = \pi\left(n + \frac{1}{2}\right), \tag{2}$$

where $(x_0, x_1)$ are the classical turning points, defined by $V(x_i) = E_n$ and $n \in \mathbb{N}_0$.

For the class of potentials we want to study, it was shown that a generalization of the Bohr–Sommerfeld rule to so-called quasi-stationary states is possible [8]. It takes the following form

$$\int_{x_0}^{x_1} \sqrt{E_n - V(x)}\, dx = \pi\left(n + \frac{1}{2}\right) - \frac{1}{2}\varphi(a). \tag{3}$$

The additional term $\varphi(a)$ involves a combination of Gamma functions

$$\varphi(a) = a(1 - \ln(a)) + \frac{1}{2i}\ln\left(\frac{\Gamma(1/2 + ia)}{\Gamma(1/2 - ia)[1 + \exp(-2\pi a)]}\right), \tag{4}$$

where

$$a = \frac{1}{\pi}\int_{x_1}^{x_2}(-p^2)^{1/2}\, dx, \qquad p = \sqrt{E_n - V(x)}. \tag{5}$$

An expansion for large $a$, which corresponds to large potential barriers, shows that equation (3) can be written in a much simpler form [7, 8]

$$\int_{x_0}^{x_1} \sqrt{E_n - V(x)}\, dx = \pi\left(n + \frac{1}{2}\right) - \frac{i}{4}\exp\left(2i\int_{x_1}^{x_2}\sqrt{E_n - V(x)}\, dx\right). \tag{6}$$

Large values of $a$ imply that the contribution of the new term in equation (6) only yields an exponentially small imaginary part. This can be used for a further expansion in equation (6) in terms of $E_n \equiv E_{0n} + iE_{1n}$, with $E_{1n} \ll E_{0n}$. From this one recovers the classical Bohr–Sommerfeld rule

$$\int_{x_0(E_{0n})}^{x_1(E_{0n})} \sqrt{E_{0n} - V(x)}\, dx = \pi\left(n + \frac{1}{2}\right), \tag{7}$$

together with the well known Gamow formula

$$E_{1n} = -\frac{1}{2}\exp\left(2i\int_{x_1(E_{0n})}^{x_2(E_{0n})}\sqrt{E_{0n} - V(x)}\, dx\right)\left(\int_{x_0(E_{0n})}^{x_1(E_{0n})}\frac{1}{\sqrt{E_{0n} - V(x)}}\, dx\right)^{-1}. \tag{8}$$

Note that this result is with respect to the real part $E_{0n}$. In the following sections we will use $E$ to describe the continuous[1] real part of the energy. The two equations (7) and (8) have been expected and they will provide us with key relations to solve the inverse problem in the next step. This technique was tested successfully in [9] for the study of oscillation spectra of ultra compact constant density stars and gravastars, demonstrating that it is an easy and reliable way to study the spectra of horizonless objects in general relativity.

---

[1] Continuous with respect to $n$, defined from the classical Bohr–Sommerfeld rule equation (7).

## 3. Inverse problem

In order to solve the inverse problem for quasi-stationary states, we will review two methods for potential wells and barriers. Afterwards we summarize our results for the generalized problem that have first been reported in [5].

### 3.1. Potential well and barrier

To reconstruct a single potential well from the knowledge of its spectrum, one can 'invert' the classical Bohr–Sommerfeld equation (2), but this does not yield a unique solution. Instead one finds the width $\mathcal{L}_1(E)$ of the potential well as a continuous function of $E$ [1, 3]

$$\mathcal{L}_1(E) \equiv x_1(E) - x_0(E) = \frac{\partial}{\partial E} I(E), \tag{9}$$

where $I(E)$ is the so-called inclusion

$$I(E) = 2 \int_{E_{\min}}^{E} \frac{n(E') + 1/2}{\sqrt{E - E'}} dE'. \tag{10}$$

$E_{\min}$ is the minimum of the potential and in general not the lowest state of the spectrum. It has to be provided as additional information or approximated from the spectrum, following from $n(E_{\min}) + 1/2 = 0$ [1, 3]. Applications of the method can be found in [10–12].

A similar relation can be found for the width $\mathcal{L}_2(E)$ of the potential barrier. It follows from 'inverting' the Gamow formula equation (8) as shown in [2, 4]

$$\mathcal{L}_2(E) \equiv x_2(E) - x_1(E) = \frac{1}{\pi} \int_{E}^{E_{\max}} \frac{(dT(E')/dE')}{T(E')\sqrt{E' - E}} dE'. \tag{11}$$

Here $E_{\max}$ is the maximum of the potential barrier and can be obtained from $T(E_{\max}) = 1$, where $T(E)$ is the so-called transmission that in the semi-classical approximation is given by

$$T(E) = \exp\left(2i \int_{x_1}^{x_2} \sqrt{E - V(X)} \, dx\right). \tag{12}$$

Note that the transmission $T(E)$ is in general not in a straightforward manner related to the spectrum $E_n$. In order to solve the inverse problem for quasi-stationary states only from the knowledge of the spectrum $E_n$, one has to connect it somehow with the transmission $T(E)$. This will be done in the next step.

### 3.2. Inverse problem for quasi-stationary states

With the results that have been presented in the previous sections, we can now generalize the inverse problem to potentials admitting quasi-stationary states.

From the expanded form of the generalized Bohr–Sommerfeld rule equation (6) it is known that the real part $E_{0n}$ is approximatively given by the classical Bohr–Sommerfeld rule equation (7), while its imaginary part $E_{1n}$ follows from the Gamow formula equation (8). The calculation of $E_{0n}$ is independent of $E_{1n}$, but not the other way round. To obtain $E_{1n}$ one has to know $E_{0n}$ first.

The reconstruction of the potential well width $\mathcal{L}_1(E)$ from the knowledge of $E_{0n}$ is straightforward by applying equation (9) for the real part of $E_n$.

A closer look to equations (8) and (12) reveals that the transmission $T(E)$ can be expressed in terms of the imaginary part of the spectrum $E_{1n}$ and an integral. The integral contains $E_{0n}$ and the knowledge of the potential $V(x)$ between $(x_0, x_1)$. From the reconstructed width $\mathcal{L}_1(E)$ one can not find a unique solution for the potential without providing one of the two turning point functions. Fortunately, it can be shown that within the Bohr–Sommerfeld rule, the result of the integral is the same for all potentials that can be constructed from $\mathcal{L}_1(E)$, as shown in the appendix in [5]. This important result is necessary to actually relate the spectrum $E_n$ with the transmission $T(E)$ and allows now to calculate $\mathcal{L}_2(E)$ from equation (11)

$$T(E) = -2E_{1n} \int_{x_0}^{x_1} \frac{1}{\sqrt{E - V(x)}} dx. \tag{13}$$

With the knowledge of $\mathcal{L}_1(E) = x_1(E) - x_0(E)$ and $\mathcal{L}_2(E) = x_2(E) - x_1(E)$ there are two equations for three turning point functions. The situation is similar to the case of potential wells and barriers, which have two turning points but only one equation to determine them. In our case, one of the three functions for the turning points has to be provided in order to find a unique solution for the potential. The potential follows from inverting the turning point functions for $E$. A more comprehensive discussion of how this problem can be addressed, by assuming that additional information about the spectrum is known, can be found in [1, 3]. An explicit example where the knowledge of the third turning point is already known from the type of problem is discussed in the subsequent section 4.

## 4. Applications

In this section we present analytic examples for the reconstruction of potentials by starting from a given function for the spectrum and analyze whether it leads to valid solutions for the potential or not. The first example is discussed in detail to demonstrate each step of the method. The subsequent examples follow the same procedure, thus only the results are summarized and discussed. The application of the method in the context of gravitational wave physics is presented in the end of this section.

### 4.1. Spectrum I

To demonstrate the method we will start from a simple analytic function for the spectrum given by

$$E_n = a\left(n + \frac{1}{2}\right) - ib \exp\left(-\frac{c}{n + 1/2}\right), \qquad (14)$$

where $(a, b, c)$ are three parameters that will be discussed later. We will assume that equation (14) describes a possible spectrum for quasi-stationary states, at least for $E_{0n} \in [E_{\min}, E_{\max}]$. It is not known to us as an exact solution for the spectrum of any potential. The inverse method allows now to find approximatively the corresponding potential. The real part of equation (14) is described by the well known harmonic oscillator spectrum, while the imaginary part is exponentially small. Such a behavior can be expected for quasi-stationary states with $E_{0n} \in [E_{\min}, E_{\max}]$. Now we want to construct the widths $\mathcal{L}_1(E)$ and $\mathcal{L}_2(E)$. Afterwards we discuss the possible potentials that can be found from them.

First we calculate $\mathcal{L}_1(E)$. For this one needs to invert the real part of the spectrum

$$n(E) = \frac{E}{a} - \frac{1}{2}. \qquad (15)$$

The minimum of the potential is extrapolated from $n(E_{\min}) + 1/2 = 0$ to be $E_{\min} = 0$. From this we can compute $\mathcal{L}_1(E)$ by using equations (9) and (10)

$$\mathcal{L}_1(E) = 2\frac{\partial}{\partial E}\int_{E_{\min}}^{E} \frac{n(E') + 1/2}{\sqrt{E - E'}} dE' = \frac{4}{a}\sqrt{E}. \qquad (16)$$

To do the calculations for $\mathcal{L}_2$ in the next step, one has to know the potential between $(x_0, x_1)$. As it is shown in the appendix of [5], we can use any of the valid potentials constructed by $\mathcal{L}_1$ without loss of generality. Let us take the symmetric potential $V_s(x)$ with $x_{\min} = 0$ by inverting the left and right symmetric turning points $(x_{s0}(E), x_{s1}(E))$

$$x_{s0}(E) \equiv -\frac{\mathcal{L}_1(E)}{2}, \qquad x_{s1}(E) \equiv \frac{\mathcal{L}_1(E)}{2}, \qquad (17)$$

from which we get

$$V_s(x) = \frac{a^2}{4}x^2. \qquad (18)$$

This recovers, as might be expected from the form of the spectrum, the harmonic oscillator potential. Nevertheless, it is worth to mention that this is only one specific solution for the potential well region. There are infinitely many other shifted or tilted potentials sharing the same $\mathcal{L}_1(E)$ and therefore the same spectrum, see [10–12] for a discussion of so-called WKB equivalent potentials.

In the next step we calculate $\mathcal{L}_2(E)$. For this, the following integral is needed

$$\int_{x_{s0}(E)}^{x_{s1}(E)} \frac{1}{\sqrt{E - V_s(x)}} dx = \frac{2}{a}\int_{-1}^{+1} \frac{1}{\sqrt{1 - u^2}} du = \frac{2\pi}{a}. \qquad (19)$$

Now we can relate the transmission $T(E)$ with the imaginary part $E_{1n}$ of the spectrum by using equations (13), (15) and (19)

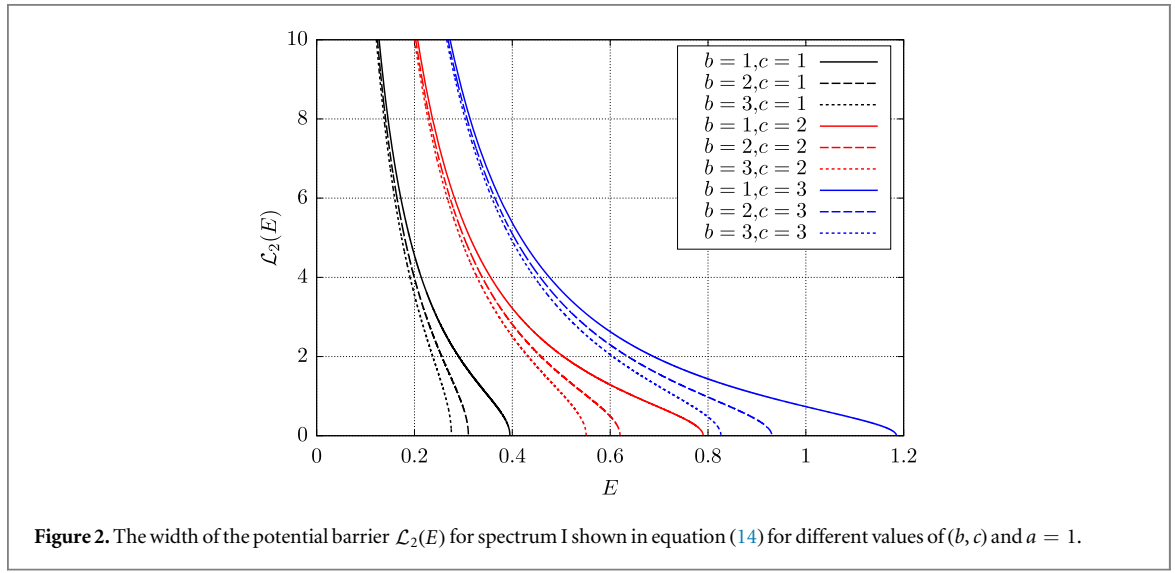

**Figure 2.** The width of the potential barrier $\mathcal{L}_2(E)$ for spectrum I shown in equation (14) for different values of $(b, c)$ and $a = 1$.

$$T(E) = -2E_{1n}\int_{x_{s0}(E)}^{x_{s1}(E)} \frac{1}{\sqrt{E - V_s(x)}}dx = -\frac{4\pi}{a}E_{1n} = \frac{4\pi b}{a}\exp\left(-\frac{c}{n(E) + 1/2}\right) = \frac{4\pi b}{a}\exp\left(-\frac{ac}{E}\right). \quad (20)$$

Using this result for the transmission in equation (11) we can finally calculate $\mathcal{L}_2(E)$

$$\mathcal{L}_2(E) = \frac{1}{\pi}\int_E^{E_{\max}} \frac{(dT(E')/dE')}{T(E')\sqrt{E' - E}}dE' = \frac{ac}{\pi}\frac{1}{E_{\max}E^{3/2}}\left(\sqrt{E_{\max} - E}\sqrt{E} + E_{\max}\tan^{-1}\left(\frac{\sqrt{E_{\max} - E}}{\sqrt{E}}\right)\right). \quad (21)$$

The only unknown left is $E_{\max}$. It follows from $T(E_{\max}) = 1$ and takes the simple form

$$E_{\max} = \frac{ac}{\ln(4\pi b/a)}, \quad (22)$$

where we have used equation (20). The parameters are not independent from each other, if they describe a spectrum related to our type of potentials. A necessary condition for them to be a valid solution is that $E_{\max}$ has to be larger than $E_{\min}$. From this one finds (for positive $c$)

$$a < 4\pi b. \quad (23)$$

The number of states between $E_{\min}$ and $E_{\max}$ can be approximated from

$$n(E_{\max}) = \frac{c}{\ln(4\pi b/a)} - \frac{1}{2}. \quad (24)$$

We show $\mathcal{L}_2$ for different parameters in figure 2. It is interesting to see that different values of $b$ have only little impact on $\mathcal{L}_2$ for small values of $E$, but become important when $E$ approaches $E_{\max}$. Changing the parameter $c$ strongly influences $\mathcal{L}_2$ for all values of $E$. The way how $E_{\max}$ depends on $(a, b, c)$ can be seen directly from equation (22) and defines the intersection of $\mathcal{L}_2 = 0$ with the $E$-axis.

After reconstructing $\mathcal{L}_1$ and $\mathcal{L}_2$ we are left with choosing one of the three turning points in order to have a unique potential. To avoid any 'overhanging cliffs' in the potential, $\mathcal{L}_1$ and $\mathcal{L}_2$, as well as the turning points $(x_0(E), x_1(E), x_2(E))$, have to be strictly monotonically increasing/decreasing functions of $E$. This has to be taken into account when choosing one of the three turning points to obtain a specific potential.

To study a wide, but not complete, range of potentials $V_\chi(x)$, we can parameterize the relation between $(x_0(E), x_1(E))$ and $\mathcal{L}_1$ in the following way

$$x_0(E) = -\chi\mathcal{L}_1(E), \qquad x_1(E) = (1 - \chi)\mathcal{L}_1(E), \quad (25)$$

with $\chi \in [0, 1]$ describing in a simple way how much the potential well is 'tilted'. The symmetric potential is recovered for $\chi = 0.5$. For all values of $\chi$ it describes a piecewise potential made of two in general different parabolas. Other parameterizations that result in more complicated shapes of the potential, e.g. around the potential barrier or $\chi = \chi(E)$, are possible. The turning point $x_2(E)$ is given by

$$x_2(E) = \mathcal{L}_2(E) + (1 - \chi)\mathcal{L}_1(E), \quad (26)$$

where we have used $\mathcal{L}_2(E) = x_2(E) - x_1(E)$ and equation (25).

From the knowledge of the turning points we can now construct $V_\chi(x)$ by solving them for $E$. We show the reconstructed potential for different values of $\chi$ in figure 3 and a fixed choice of the parameters $(a, b, c)$.

With our choice for the parameterization, the turning points $(x_0(E), x_1(E))$ are by construction strictly monotonically decreasing and increasing functions of $E$, respectively. We can now determine what choices of $\chi$

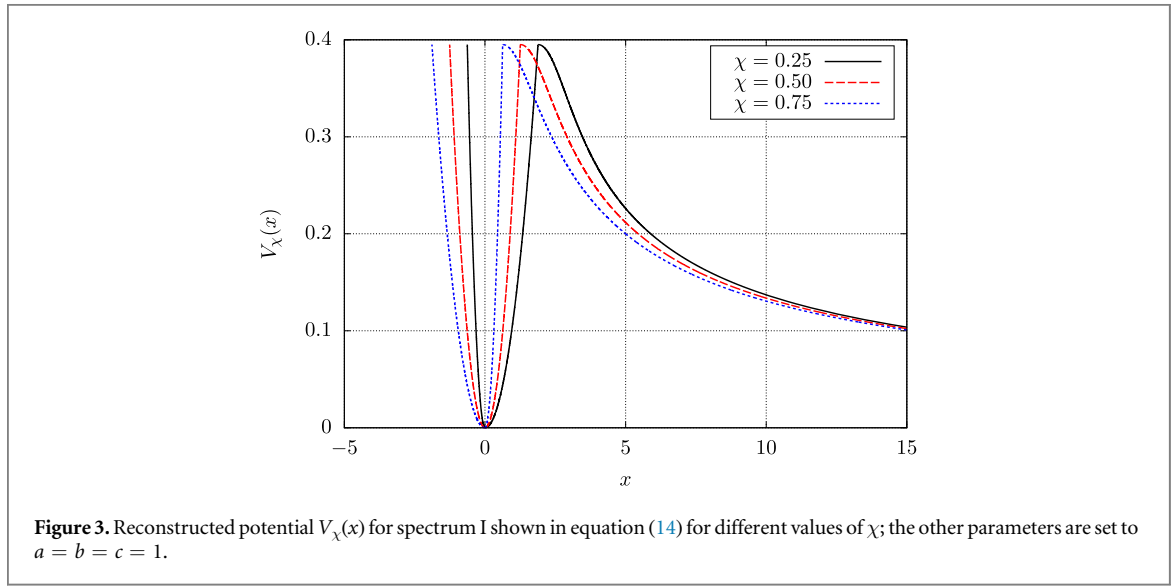

**Figure 3.** Reconstructed potential $V_\chi(x)$ for spectrum I shown in equation (14) for different values of $\chi$; the other parameters are set to $a = b = c = 1$.

are valid to guarantee that $x_2(E)$ is strictly monotonically decreasing by demanding

$$\frac{d}{dE}x_2(E; \chi, a, b, c) < 0, \qquad \forall E \in (E_{\min}, E_{\max}). \tag{27}$$

This algebraic relation is too involved to be solved easily as a general function of $(\chi, a, b, c)$, but can be checked numerically for a given choice of the parameters.

### 4.2. Spectrum II
In this example we keep the same real part as in the first case but change the imaginary part

$$E_n = a\left(n + \frac{1}{2}\right) - ib \exp\left(-\frac{c}{\sqrt{n + 1/2}}\right). \tag{28}$$

Since the real part is the same, $\mathcal{L}_1(E)$ is given by equation (16), but $\mathcal{L}_2(E)$ will be different because the transmission $T(E)$ takes another form

$$T(E) = \frac{4\pi b}{a}\exp\left(-\frac{c\sqrt{a}}{\sqrt{E}}\right). \tag{29}$$

The integration of equation (11) to obtain $\mathcal{L}_2(E)$ is straightforward and leads to

$$\mathcal{L}_2(E) = \frac{c\sqrt{a}}{\pi}\frac{\sqrt{1 - E/E_{\max}}}{E}, \tag{30}$$

where $E_{\max}$ is determined by $T(E_{\max}) = 1$ and given by

$$E_{\max} = \frac{ac^2}{[\ln(4\pi b/a)]^2}. \tag{31}$$

We show $\mathcal{L}_2(E)$ for different parameters in figure 4. One can check that $\mathcal{L}_2(E)$ describes a valid potential barrier by looking at the condition

$$\frac{d\mathcal{L}_2(E)}{dE} = \frac{c\sqrt{a}}{2\pi}\frac{E - 2E_{\max}}{E^2 E_{\max}\sqrt{1 - E/E_{\max}}} < 0, \qquad \forall E \in (E_{\min}, E_{\max}). \tag{32}$$

This condition is always fulfilled (assuming $a > 0$), therefore $\mathcal{L}_2(E)$ is a valid solution to describe the width of the potential barrier. Using the same parameterization as for the first spectrum, we show the reconstructed potential $V_\chi(x)$ for a given choice of $(a, b, c)$ in figure 5.

### 4.3. Spectrum III
The third example shall be given by

$$E_n = a\left(n + \frac{1}{2}\right) - i\exp\left(-\left[N - b\left(n + \frac{1}{2}\right)\right]\right), \tag{33}$$

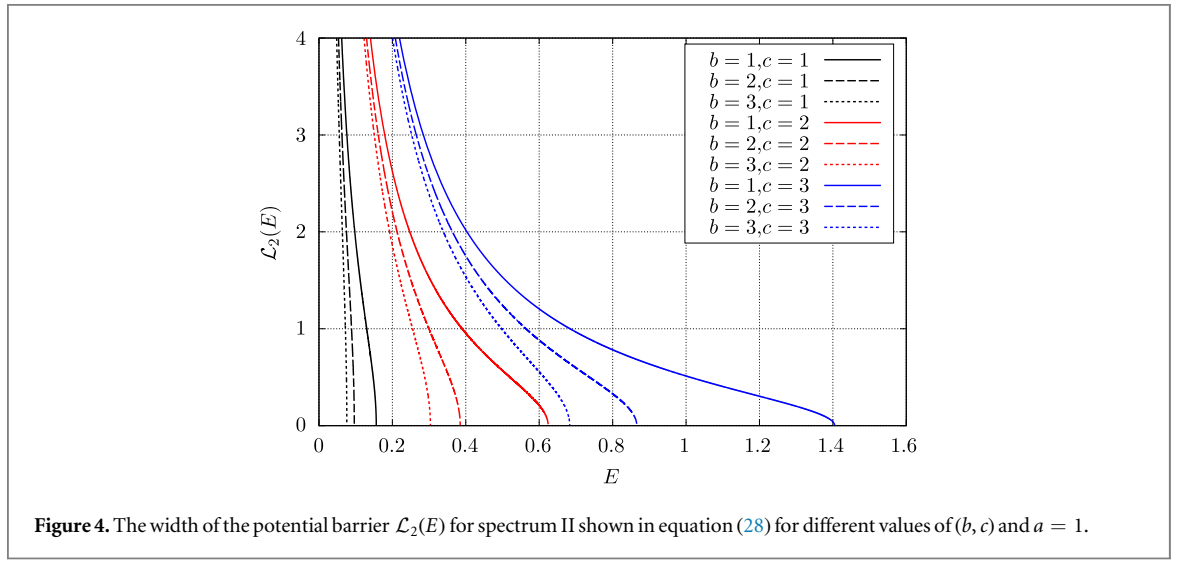

**Figure 4.** The width of the potential barrier $\mathcal{L}_2(E)$ for spectrum II shown in equation (28) for different values of $(b, c)$ and $a = 1$.

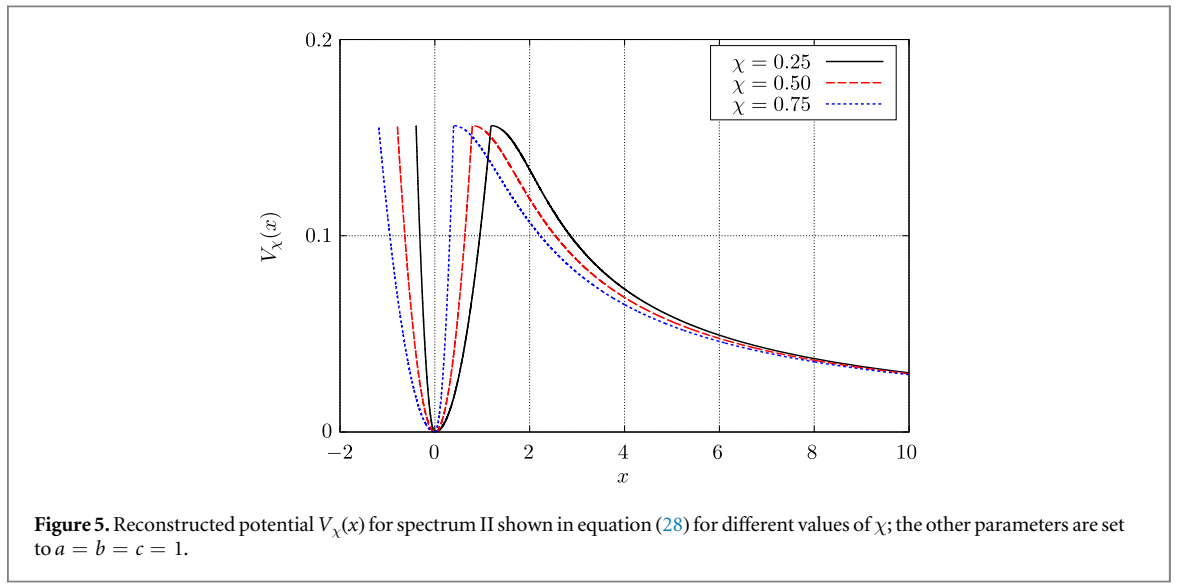

**Figure 5.** Reconstructed potential $V_\chi(x)$ for spectrum II shown in equation (28) for different values of $\chi$; the other parameters are set to $a = b = c = 1$.

using the parameters $(a, b, N)$. Following the same procedure as in the previous examples one finds

$$T(E) = \frac{4\pi}{a}\exp\left(-N + \frac{b}{a}E\right), \quad \mathcal{L}_2(E) = \frac{2b}{\pi a}\sqrt{E_{\max} - E}, \quad E_{\max} = \frac{a}{b}\left[N + \ln\left(\frac{a}{4\pi}\right)\right]. \qquad (34)$$

The special case of a symmetric potential barrier reveals that it is described by an inverted parabola with the maximum at $E_{\max}$. We show $\mathcal{L}_2(E)$ in figure 6 for $a = 1$ and different values of $(b, N)$.

Both functions, $\mathcal{L}_1(E)$ and $\mathcal{L}_2(E)$, are strictly monotonically increasing and decreasing functions of $E$ for $E \in (E_{\min}, E_{\max})$, respectively. Nevertheless, if we choose to parameterize a subclass of the possible solutions for the potentials as done in the first example, something interesting happens. Because overhanging cliffs in the potential are not allowed, a quick analysis shows that only $\chi = 1$ gives a valid solution. For all other values of $\chi$ one finds overhanging cliffs for $E \to E_{\min}$. This can be seen by using equation (26) to define $x_2(E)$, which has to be a strictly monotonically decreasing function of $E$, and check whether $dx_2(E)/dE$ changes sign. One finds that this is the case at a given value $E_c$, which is given by

$$E_c = \frac{E_{\max}}{1 + [b/(2\pi(1 - \chi))]^2}. \qquad (35)$$

The parameter $\chi$ is defined in the interval $[0, 1]$. For $\chi \neq 1$ we therefore always find a value of $E_c$ in the range $(E_{\min}, E_{\max})$. This excludes all potentials described by $\chi \neq 1$ from being valid solutions. The special case of $\chi = 1$ is shown in figure 7.

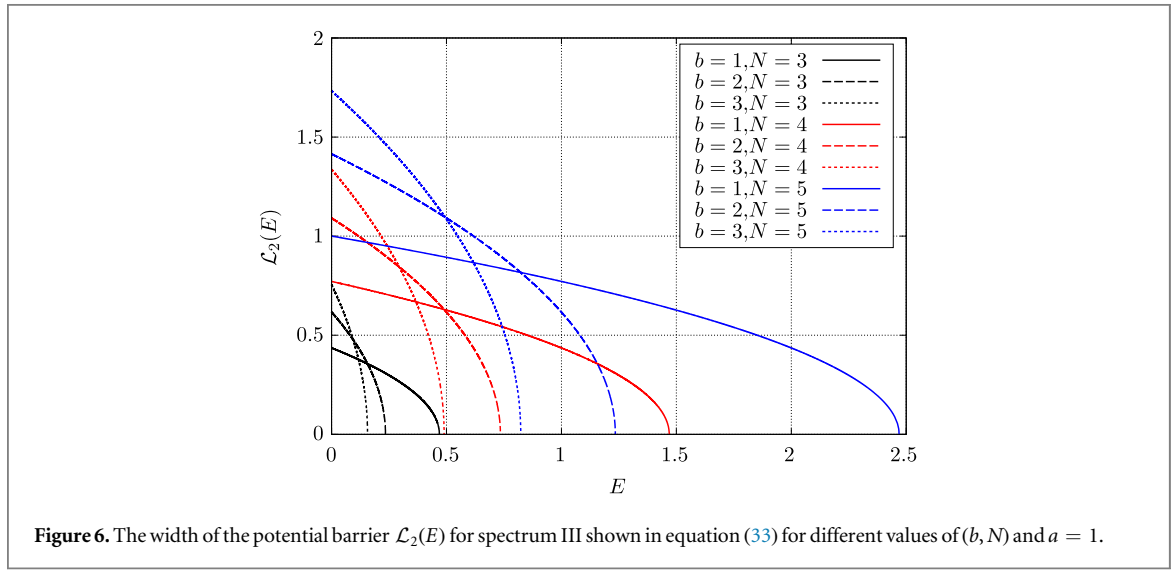

**Figure 6.** The width of the potential barrier $\mathcal{L}_2(E)$ for spectrum III shown in equation (33) for different values of $(b, N)$ and $a = 1$.

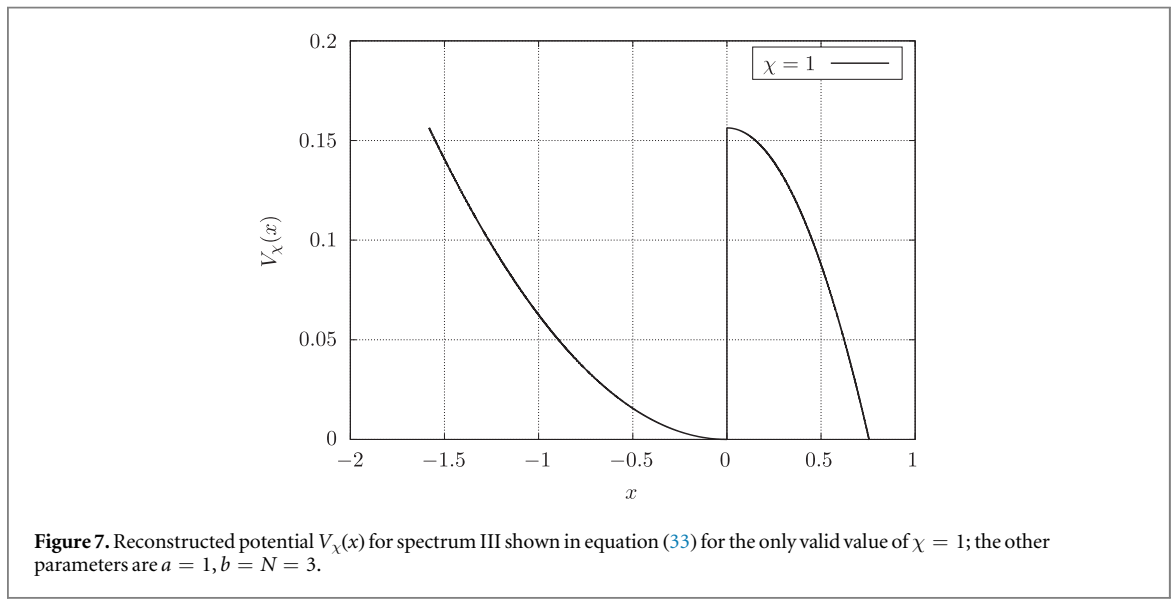

**Figure 7.** Reconstructed potential $V_\chi(x)$ for spectrum III shown in equation (33) for the only valid value of $\chi = 1$; the other parameters are $a = 1, b = N = 3$.

### 4.4. Spectrum IV

In this last example we show that not every function for the spectrum leads to a valid solution for the potential. This is the case for the following spectrum

$$E_n = a\left(n + \frac{1}{2}\right) - i\exp\left(-\left[N - b\left(n + \frac{1}{2}\right)^2\right]\right), \quad (36)$$

where we keep the same real part as in the previous examples but modified the imaginary part. The calculations are again straightforward and the important relations are found to be

$$T(E) = \frac{4\pi}{a}\exp\left(-\left[N - \frac{b}{a^2}E^2\right]\right), \quad \mathcal{L}_2(E) = \frac{4}{3\pi}\frac{b}{a^2}\sqrt{E_{\max} - E}\,(E_{\max} + 2E),$$

$$E_{\max} = \sqrt{\frac{a^2}{b}\left[N + \ln\left(\frac{a}{4\pi}\right)\right]}. \quad (37)$$

A quick look at $\mathcal{L}_2(E)$ shows that its derivative with respect to $E$ is positive for $E \in E(E_{\min}, E_{\max}/2)$, which would cause overhanging cliffs for the potential barrier and is therefore clearly no valid solution for a potential. In this case we conclude that there exists no valid potential in our class of potentials that are described by the inverse method, which can have the given spectrum equation (36). Nevertheless, we show the corresponding $\mathcal{L}_2(E)$ for $a = 1$ and different values of $(b, N)$ in figure 8.

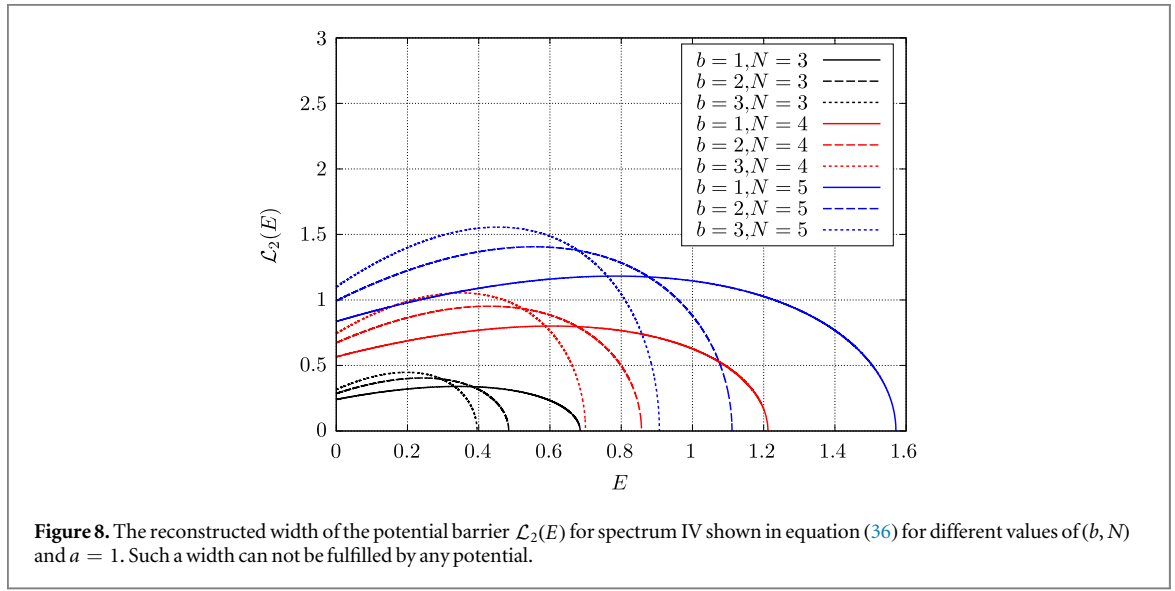

**Figure 8.** The reconstructed width of the potential barrier $\mathcal{L}_2(E)$ for spectrum IV shown in equation (36) for different values of $(b, N)$ and $a = 1$. Such a width can not be fulfilled by any potential.

### 4.5. Gravitational waves

The ground-breaking measurements of gravitational waves emitted from the merger of two black holes and more recently from the merger of two neutron stars, by the gravitational wave detectors advanced LIGO and Virgo [13–17], open new possibilities to study compact relativistic objects. Like in the spectroscopy of atoms, a rich zoo of various kind of modes and perturbations appears that can potentially be measured and used to study the equation of state of matter at extreme densities or to test the general theory of relativity in the so-called strong-field regime. For a broad class of hypothetical ultra compact objects, the relevant problem can in some cases be simplified to the study of an one-dimensional wave equation with potentials of the type discussed in this work. For such objects it was shown that the expected gravitational wave signal could actually mimic black holes at early times of a signal, but features a series of so-called 'echoes' as distinct feature [18, 19]. The interest in so-called exotic compact objects increased tremendously after tentative evidence in the binary black hole merger detections has been found [20] and recently potentially confirmed [21]. However, due to the weak signal, technical difficulties concerning the involved data analysis and physical models being used, the correct interpretation of the signal and validity of the physical models under consideration are highly debated in the literature [22–24].

The inverse method presented here has recently been applied in the study of gravitational perturbations of ultra compact objects within general relativity [5]. The function for the third turning point can naturally be provided by Birkhoff's theorem, which states that the space-time outside spherically symmetric and non-rotating matter configurations has to be the Schwarzschild space-time. Using this additional information it is possible to obtain a unique, but approximate reconstruction of the so-called perturbation potential of the axial modes that characterize the object.

In this work, the spectrum was assumed to be known as a finite and discrete set of complex numbers. It was provided by a numerical code presented in [25], which solves the direct problem for different objects. The lack of the analytic form of the spectrum made it necessary to use inter-/extrapolation techniques to have continuous functions of the spectrum and the transmission for the integration procedure. It was found that the agreement of the true potential with the reconstructed one improves significantly for cases where the number of states within the potential well is large. This is because the underlying semi-classical methods can be expected to work more precisely for large values of $n$ and the inter-/extrapolation to approximate ($E_{\min}$, $E_{\max}$) become more precise then as well. The application of the method to observed gravitational wave data is more involved and requires additional knowledge for the interpretation and analysis of such signals, as well as events with a higher signal to noise ratio. First steps in this direction have already been undertaken in [26–29].

## 5. Discussion

We have applied the inverse method to four different analytic functions that were assumed to be valid spectra for quasi-stationary states in a potential of the type that is studied. For the part of the spectrum being associated with a potential region between $E_{\min}$ and $E_{\max}$, the spectrum should have a much larger real than imaginary part. Both, the real and imaginary part, have to grow with increasing $n$. We have taken this into account by choosing functions with exponentially small imaginary part, which is expected from the Gamow formula. Even if this qualitative criterion is fulfilled, we found that not all such functions lead to valid potentials, see spectrum 4 in 4.4. All of the studied spectra have three parameters that were not specified explicitly in the beginning. However, in order to obtain consistent potentials, the parameters are in general not independent from each other and have to fulfill certain conditions, e.g. $E_{\max} > E_{\min}$ or $d\mathcal{L}_2(E)/dE < 0$.

Other interesting considerations arise if the inverse method is applied to an observed spectrum, which is known as discrete set of complex numbers that also includes errors. Unfortunately a straightforward error calculation is not possible, due to the intrinsically approximate character of the underlying WKB method and the complicated dependency of the spectrum in the integral equations that have to be solved. Nevertheless, we had a look at this issue and want to make some qualitative comments that can be valuable if the method is applied to such cases. The key observation is that the errors in the reconstructed widths can strongly depend on the way how the continuous spectrum is inter-/extrapolated from the discrete spectrum and what type of errors are assumed. For example, adding alternating errors to the spectrum would cause larger errors if one uses higher order splines in the inter-/extrapolation, while the results are much more stable if the spline order is low and the errors are random or smoothed. Another important consideration is that the functions being constructed in this way for the spectrum $n(E)$ and the transmission $T(E)$ can not be arbitrary. They must fulfill a few conditions to bear a consistent meaning in the semi-classical description. These conditions have to be considered to verify that the inter-/extrapolation yields physical results, e.g. both functions must be strictly monotonically increasing functions of $E$. Also the reconstructed widths $\mathcal{L}_1(E)$ and $\mathcal{L}_2(E)$ must be strictly monotonically increasing and decreasing functions of $E$, respectively.

As a last remark, we want to mention that even if the reconstructed potential has a consistent shape, one should in principle also check whether there is a reasonable number of quasi-stationary states for the actual choice of parameters. The number of states can be approximated from $n(E_{\max})$. In the provided figures we want to show the qualitative properties of $\mathcal{L}_2(E)$ in a clearly arranged way. Therefore we have decided to use a simple sample of numerical values for the parameters. This choice does not necessarily correspond to many states in the potential, but its qualitative shape is not affected by this. A set of parameters giving a large number of states can in general be found by demanding large $n(E_{\max})$ and using the provided necessary conditions.

## 6. Conclusions

In this work we have presented a simple method to solve the inverse spectrum problem for a specific class of potentials that admit quasi-stationary states. The method is based on a generalization of semi-classical methods for the inverse problem of potential wells and potential barriers [1–4]. It was shown how the knowledge of the complex spectrum $E_n$ can be used in multiple steps to reconstruct the widths of the potential well $\mathcal{L}_1(E) = x_1(E) - x_0(E)$ and the potential barrier $\mathcal{L}_2(E) = x_2(E) - x_1(E)$, where $(x_0, x_1, x_2)$ are the turning points. As already known from similar problems, it turns out that the reconstruction is not unique, unless a function for one of the three classical turning points is known. In practice this might naturally be provided by additional available information. We discuss such a case in the recent application of the presented method in gravitational wave physics [5]. To demonstrate the method, we have solved the inverse problem for four analytic spectra and investigated the properties of the solutions, as well as the parameter space.

Since the method allows for a simple reconstruction of quasi-stationary state potentials from a given spectrum in the one-dimensional wave equation, it might be interesting for different branches of physics and beyond.

## Acknowledgments

S H Völkel wants to thank A Boden for useful discussions as well as K D Kokkotas, D D Doneva, P Pnigouras and A Maselli for valuable feedback that improved the final version of this work. The constructive reports of the anonymous referees are acknowledged with many thanks. The author is indebted to the Baden-Württemberg Stiftung for the financial support of this research project by the Eliteprogramme for Postdocs.

## ORCID iDs

Sebastian H Völkel  https://orcid.org/0000-0002-9432-7690